\documentclass[twoside,journal]{IEEEtran}

\usepackage{makecell}
\usepackage{array}
\usepackage{graphicx,amssymb,amsmath}
\usepackage{multicol}
\usepackage[noadjust]{cite}
\usepackage{setspace}
\usepackage{subfigure}
\usepackage{graphicx}
\usepackage{float}
\usepackage {url}
\usepackage{stfloats}
\usepackage{amsthm,pifont}
\usepackage{flushend}
\usepackage{cases,subeqnarray}
\usepackage{bm,multirow,bigstrut}
\usepackage{amsmath, amsthm, amssymb}
\usepackage{textcomp}
\usepackage{latexsym,bm}
\usepackage{booktabs}
\usepackage{xcolor}
\usepackage{mathtools}
\usepackage{dsfont}
\usepackage{extarrows}
\usepackage{epsfig}
\usepackage{epsfig}
\usepackage{epstopdf}
\usepackage[noend]{algpseudocode}
\usepackage{algorithmicx,algorithm}

\usepackage{cite}
\usepackage{bm}
\usepackage{multicol}       
\usepackage{multirow}       
\usepackage{array}          
\usepackage{colortbl}
\usepackage{makecell}
\definecolor{crimson}{RGB}{192,0,0}         
\definecolor{navy}{RGB}{47,85,151}         
\usepackage{bbding}
\usepackage{graphicx}
\usepackage{booktabs}
\usepackage{algorithm}
\usepackage{algpseudocode}
\usepackage{diagbox}

\usepackage{graphicx}
\usepackage{setspace}
\usepackage{cleveref}

\usepackage{url}
\usepackage{mathtools}
\def\BibTeX{{\rm B\kern-.05em{\sc i\kern-.025em b}\kern-.08em
 T\kern-.1667em\lower.7ex\hbox{E}\kern-.125emX}}

\usepackage{stackengine}
\usepackage{enumerate}

\usepackage{multicol}
\usepackage[noend]{algpseudocode}
\usepackage{pgfplots}
\makeatletter
\def\BState{\State\hskip-\ALG@thistlm}
\makeatother
\usepackage{multirow}
\definecolor{Blues}{RGB}{0,0,0}
\definecolor{mygray}{gray}{.8}
\definecolor{mygray2}{gray}{.7}
\definecolor{mygray3}{gray}{.6}

\algdef{S}[FOR]{ForEach}[1]{\algorithmicforeach\ #1\ \algorithmicdo}

\theoremstyle{plain}

\theoremstyle{plain}

\usepackage{amsmath}

\IEEEoverridecommandlockouts

\begin{document}

\title{Flexible MIMO for Future Wireless Communications: Which Flexibilities are Possible?}
\author{Zhe Wang, Jiayi Zhang,~\IEEEmembership{Senior Member,~IEEE}, Bokai Xu, Wenhui Yi, \\Emil Bj{\"o}rnson,~\IEEEmembership{Fellow,~IEEE}, Bo Ai,~\IEEEmembership{Fellow,~IEEE}\vspace*{-0.3cm}
\thanks{Z. Wang, J. Zhang (corresponding author), B. Xu, W. Yi, and B. Ai are with the State Key Laboratory of Advanced Rail Autonomous Operation and the School of Electronics and Information Engineering, Beijing Jiaotong University, Beijing 100044, China. J. Zhang is also with Nanjing Rongcai Transportation Technology Research Institute Co., Ltd., Nanjing 210000, China; Z. Wang is also with the Division of Communication Systems, KTH Royal Institute of Technology, Stockholm 10044, Sweden; E. Bj{\"o}rnson is with the Division of Communication Systems, KTH Royal Institute of Technology, Stockholm 10044, Sweden.}\vspace{-0.3cm}}
\maketitle

\begin{abstract}
In conventional multiple-input multiple-output (MIMO), static array configurations struggle in dynamic environments, and further antenna scaling is bounded by cost, energy, and footprint. Emerging approaches, which can enable next-generation wireless communication networks with modest spectrum availability by leveraging flexibility and adaptability rather than sheer array growth, are therefore needed. In this paper, we present a taxonomy framework, referred to as flexible MIMO technology, that systematically categorizes a wide range of evolving MIMO technologies. The focus is on MIMO technologies with flexible physical configurations and integrated applications. We categorize twelve representative flexible MIMO technologies into three major classifications: flexible deployment characteristics-based, flexible geometry characteristics-based, and flexible real-time modifications-based. We then comprehensively overview their fundamental characteristics, potential, and challenges. In addition, we highlight three vital enablers for flexible MIMO technology, including efficient channel state information acquisition schemes, low-complexity beamforming design, and explainable artificial intelligence (AI)-enabled optimization, and discuss eight representative sub-techniques. Finally, two brief case studies---pre-optimized irregular array for high-speed railway network and cell-free movable antenna---are presented, showing how flexible MIMO can open new design possibilities and inspire future research directions for next-generation wireless networks.

\end{abstract}

\section{Introduction}
Multiple-input multiple-output (MIMO) technology has been instrumental in the evolution of wireless networks. It has achieved remarkable success in the fifth-generation (5G) wireless communication networks, relying on its ability to boost spectral efficiency (SE) through spatial multiplexing and beamforming. Looking ahead to future generations, wireless communication networks will face even more stringent requirements across an ever-wider area of application scenarios. To facilitate these requirements and application scenarios, MIMO technology needs to evolve further. On the one hand, in traditional MIMO technology, the base station (BS) is equipped with a uniform planar array (UPA) or a uniform linear array (ULA), which remains static after its deployment. However, this fixed array configuration lacks flexibility and adaptability to face dynamically varying wireless channel environments and enable future application scenarios. On the other hand, one natural evolution idea of MIMO technology is to scale up the number of antennas at the BS. However, it's not economical and sustainable to always scale up the antenna number because of cost, power consumption, device footprint, etc. Meanwhile, massive MIMO builds on the idea of having a surplus of antennas compared to users, such as 8$\times$ more, because that is needed with classical arrays to achieve favorable channel conditions. With flexible MIMO, we can obtain a more graceful antenna scaling thanks to the new features. Thus, it is essential to explore and develop next‑generation MIMO technologies that offer geometric and topological flexibility, empowering future wireless systems with the performance and versatility they require \cite{bjornson2019massive,10858129}.

Motivated by the above observations, in this paper, we propose a new overarching taxonomy framework: flexible MIMO technology. \textbf{\emph{Flexible MIMO denotes a MIMO technology with flexible physical configurations and integrated applications, including flexible deployment characteristics, geometry characteristics, and real-time modified configurations.}} In flexible MIMO technology, the network topology, integrated application, matrix geometry characteristic, and real-time antenna movement can be configured flexibly and dynamically. More specifically, the network topology denotes the network device and the network deployment strategy. The integrated application represents the services enabled by the MIMO technology. The characteristic of the array geometry includes the shape of the array, the orientation/brightness of the array, and the antenna deployment strategy. Real-time antenna movement is the further modification strategy beyond the above technologies, empowered by instantaneous channel state information (CSI). Indeed, by flexibly reconfiguring multi-antenna arrays in the above four major aspects, the flexible MIMO technology can be effectively utilized to empower future wireless communication networks. 
The purpose of introducing ``flexible MIMO" is not to coin another isolated label, but to establish a taxonomy framework for many emerging MIMO technologies. While numerous concepts have appeared, they are often presented separately and lack a holistic vision. This framework, motivated by the new MIMO flexibilities, enables wireless practitioners to adopt a generalized perspective and derive clear, systematic research clues for the future evolution of MIMO.

By utilizing the flexible MIMO technology, some major benefits can be achieved: \textbf{\emph{1) Strengthened macro and micro diversity}}: The flexible MIMO technology can significantly strengthen both the macro and micro diversity. The macro diversity can be enhanced by novel network topologies, such as the cell-free network \cite{10422885} or pinching antenna system (PASS) \cite{liu2025pinching}. The micro diversity can also be strengthened by the real-time small-scale fading-based flexible technologies, such as the real-time antenna movement-based technology \cite{10906511}. \textbf{\emph{2) Ability to reconfigure wireless channel conditions}}: Relying on flexible deployment and geometry characteristics, the wireless channel condition can be consequently reconfigured by the flexible MIMO technology, enhancing favorable wireless channel conditions for the transmission. The conditions can be optimized at different granularity, from the deployment phase to the small-scale fading level.

In this paper, we investigate a novel taxonomy framework: flexible MIMO technology to empower future wireless communication networks. The major contributions are summarized as follows.
\begin{itemize}
\item We classify and introduce twelve flexible MIMO technologies, organized into three categories: flexible deployment characteristics based on, flexible geometry characteristics based on, and flexible real-time modifications based on. Then, we introduce the fundamentals of these technologies and bring promising design motivations.
\item To facilitate the flexible MIMO technology, three promising enablers are studied, including efficient CSI acquisition schemes, low-complexity beamforming design, and explainable artificial intelligence (AI)-enabled optimization design, where eight insightful sub-enabling technologies are introduced. 
\item In case studies, we study two promising flexible MIMO technologies, pre-optimized irregular arrays for high-speed railway networks, and cell-free movable antennas. Then, we demonstrate the promising potential for flexible MIMO technologies to enhance the network capacity. 

\end{itemize}


\section{Fundamentals of Flexible MIMO Technology}


The flexibility of MIMO technology comes mainly from its physical configurations and integrated applications, including flexible deployment characteristics, geometry characteristics, and modifiable real-time configurations.

\subsubsection{Flexible Deployment Characteristics}
The deployment characteristics include mainly the network topology and integrated application scenario. The network topology refers to the configuration of multi-antenna-based wireless networks, such as the type and arrangement strategy of multi-antenna devices. Moreover, multi-antenna devices in various application scenarios can also exploit flexibility.

\subsubsection{Flexible Geometry Characteristics}
The previous classification is motivated from the network deployment level, while the second classification is from the antenna array level. Building on the flexible deployment characteristics, the geometry properties of the antenna array, including its shape, spatial position, and antenna element arrangement, can be utilized to further enhance the flexibility of wireless communication networks.

\subsubsection{Flexible Real-Time Modifications}
The above two classifications are investigated from the network and antenna array levels, respectively, which are normally configured before the real-time instantaneous information-based phase. Notably, flexible real-time modifications for the MIMO technology can be further implemented to utilize instantaneous small-scale fading effects. However, the necessary channel information for these flexible real-time modification-based technologies is much deeper than what one needs for normal MIMO communications, since we need to be able to predict the channel information not only at the current antenna locations but elsewhere.

\begin{figure*}[t]
\setlength{\abovecaptionskip}{-0.1cm}
\centering
\includegraphics[scale=0.28]{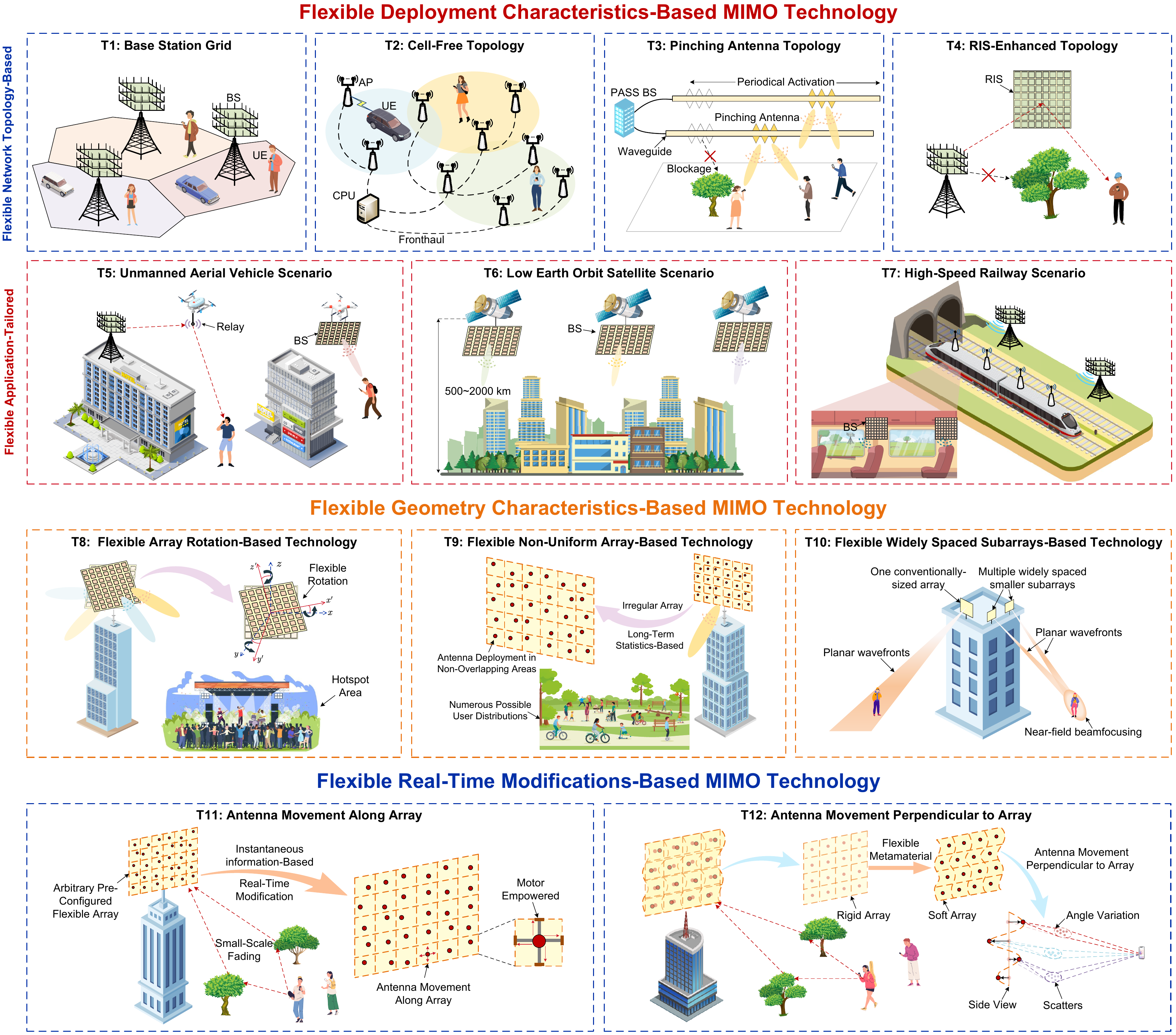}
\caption{Illustration of flexible MIMO technologies. \label{technology}}
\vspace{-0.4cm}
\end{figure*}
 
\subsection{Flexible Deployment Characteristics-Based Technology}
We will now introduce promising flexible MIMO technologies based on the above flexibility sources as shown in Fig.~\ref{technology}. The first classification is the flexible deployment characteristics-based technology, which can be further classified from the network topology and integrated application-tailored one.
\subsubsection{Flexible Network Topology-Based}\label{network_topo}
The network topology can showcase abundant flexibility with the aid of flexible configurations of base stations (BSs).

$\bullet$ \emph{T1: Base Station Grid} 

In the conventional network topology, each BS creates a cell, and user equipments (UEs) located in one particular cell are served by the corresponding BS. This cellular topology is widely applied in the massive MIMO technology and in all cellular networks of today. However, for this cellular topology, UEs at the cell edge may achieve poor service quality due to the severe inter-cell interference and significant decrease in the desired signal.

$\bullet$ \emph{T2: Cell-Free Topology}

One prominent network topology beyond the conventional cellular one is the cell-free network \cite{10422885}, where numerous BSs with small sizes and a few antennas each are deployed in a wide coverage area in a distributed manner. These APs are linked to one or several central processing units, and they perform coherence joint transmission/reception, which resolves the inter-cell interference that would otherwise be the issue in the dense cellular network \cite{10422885}.

$\bullet$ \emph{T3: Pinching Antenna Topology} 

A recent widely studied network topology, the PASS \cite{liu2025pinching}, employs a low-loss dielectric waveguide with clip-on separated dielectrics (``pinching antennas"). Unlike leaky-feeder cables used in tunnels, PASS allows the number and positions of pinched dielectrics to be adjusted by simple attach/detach operations across the waveguide. By incorporating multiple waveguides at the BS, PASS enables near-wired transmission and stable line-of-sight (LoS) links between the dielectrics and UEs. The insertion loss along the waveguide can be described using models such as \cite[Eq. (2)]{liu2025pinching}. Considering the linear feature of the waveguide, the practical outdoor deployment of PASS is one important open challenge.

$\bullet$ \emph{T4: RIS-Enhanced Topology}

Another widely studied network topology is the reconfigurable intelligent surfaces (RIS)-enhanced one \cite{enyusurvey}. In this topology, the RIS can be strategically deployed in BS-enabled wireless networks, as introduced above, to reflect and shape electromagnetic (EM) radio waves. By applying the RIS, the wireless channel environment can be flexibly reconfigured with low cost, and the coverage ability can also be enhanced.

\subsubsection{Flexible Application-Tailored}\label{application}
Multi-antenna devices in various application scenarios also showcase significant flexibility, where tailored multi-antenna devices can be applied to empower respective application scenarios.

$\bullet$ \emph{T5: Unmanned Aerial Vehicle Scenario}

By equipping UAVs with multiple antennas, they can bring a new processing dimension in the aerial domain and can serve as the transceiver or relay \cite{10422885}. Due to their high altitude, UAVs can bring LoS propagation conditions with a high probability for outdoor UEs. Besides, the cellular or cell-free networks can be effectively created using UAVs, but with the added benefit that the BS locations can be changed as the devices move. Moreover, UAVs can be flexibly deployed on demand, which is appealing for temporary or unexpected scenarios, such as disaster scenarios.

$\bullet$ \emph{T6: Low Earth Orbit Satellite Scenario}

MIMO technology can also be applied to low earth orbit (LEO) satellites to empower global wireless access \cite{9110855}. The altitude of LEO satellites typically ranges from $500\,\mathrm{km}$ to $2000\,\mathrm{km}$. The extremely large transmitting distance makes it vital to consider practical factors, such as severe pathloss and delay. The arrays in the LEO satellite are used both to achieve beamforming gains and to create spot beams on the ground, which basically create multiple cells per satellite.

$\bullet$ \emph{T7: High-Speed Railway Scenario}

High-speed railway (HSR) has become an important transportation tool. Train-mounted multi-antenna devices are tailor-designed to empower the HSR scenario \cite{10422885}. For instance, multi-antenna arrays can be deployed on the roof of the train or the wall in the carriage of the train to transmit or receive the EM signals from surrounding BSs. Notably, the high-speed movement of HSR provides challenges for the utilization of MIMO technology. However, the fixed travel route and precise scheduling can benefit the design of MIMO technology in HSR.

\begin{table*}[t]
\setlength{\abovecaptionskip}{-0.1cm}
\centering
\caption{Characteristics, potentials, and key challenges of all studied flexible MIMO technologies. \label{Comparison}}
\includegraphics[scale=0.5]{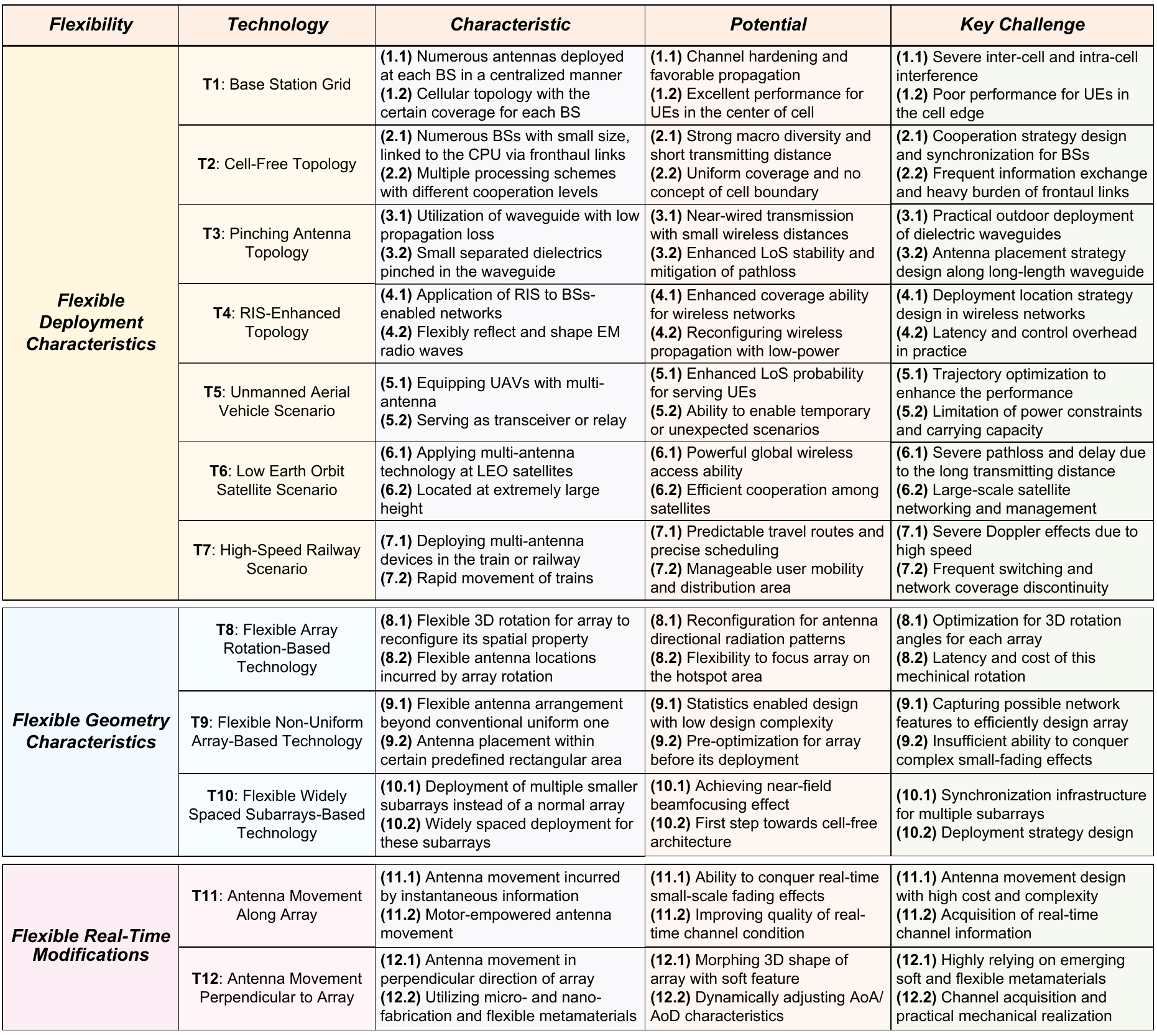}
\vspace{-0.6cm}
\end{table*}

\subsection{Flexible Geometry Characteristics-Based Technology}\label{Sec_geo}
Symmetric arrays such as UPAs and ULAs are widely used in both practice and research. While this geometry can be motivated by the sampling theorem in an isotropic propagation environment, the geometry characteristics can be optimized for specific wireless environments.

\begin{figure*}[t]
\setlength{\abovecaptionskip}{-0.1cm}
\centering
\includegraphics[scale=0.43]{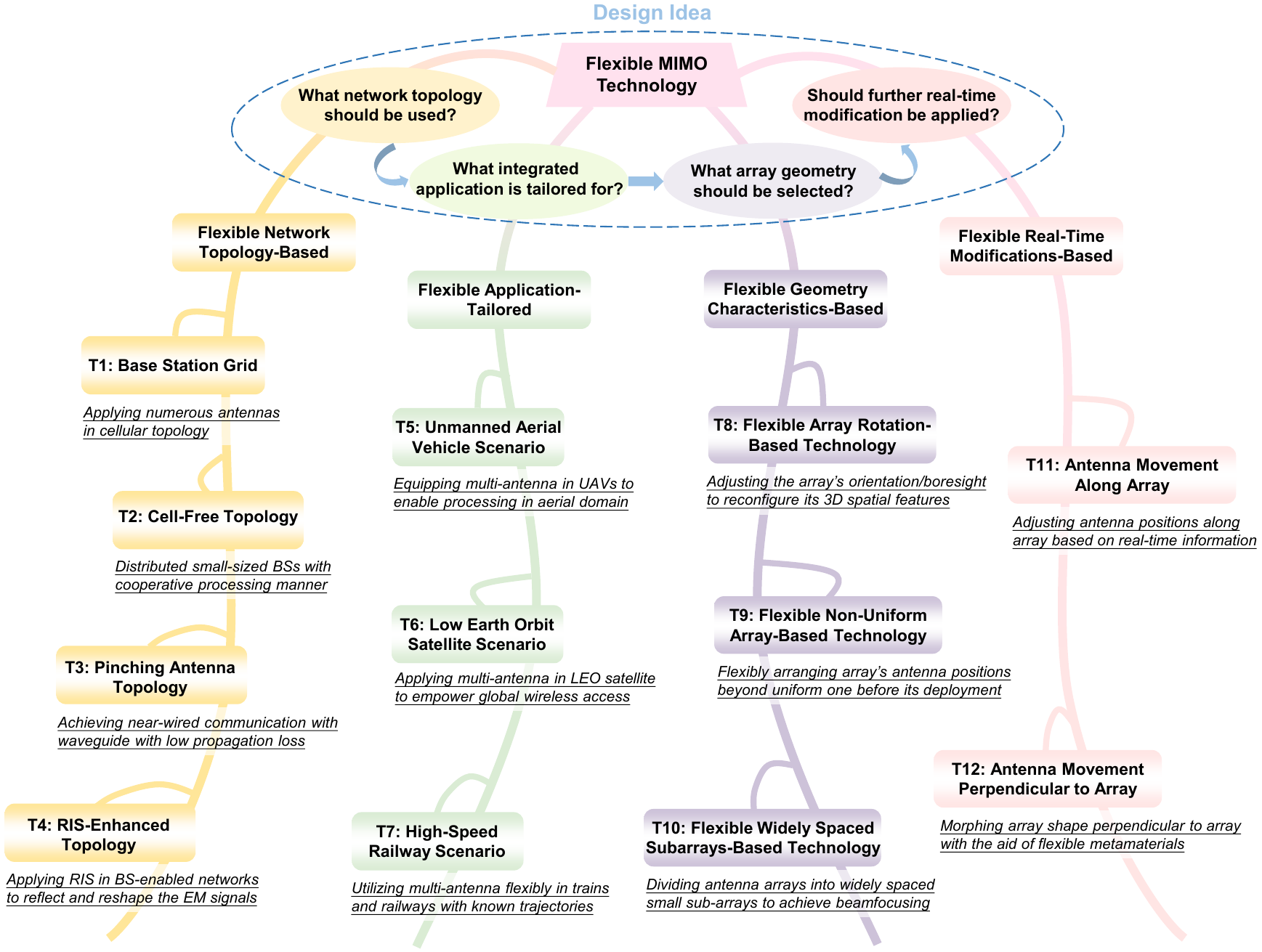}
\caption{Design idea and research clues for flexible MIMO technologies. \label{Clue_Reference}}
\vspace{-0.5cm}
\end{figure*}

$\bullet$ \emph{T8: Flexible Array Rotation-Based Technology}

The array can be flexibly rotated to exploit the potential rotation gain in the angular domain \cite{shao2025tutorial}. By adjusting the array's orientation and boresight to reconfigure its 3D spatial properties, several potential benefits can be achieved. First, the antenna directional radiation pattern can be flexibly reconfigured, which is especially useful in scenarios with directional antennas or scenarios with quite large steering angles, leading to the directivity loss for normal antennas. Second, in practice, user distributions are often unpredictable, leading to specific hotspot areas requiring high-quality service. The array aperture looks smaller in non-broadside directions, which leads to wider beams and reduced spatial resolution, but this can be circumvented through rotation. Third, polarization mismatches can be effectively mitigated by rotating the antenna array.

$\bullet$ \emph{T9: Flexible Non-Uniform Array-Based Technology}

The conventional uniform array may not always be the best topology choice. Instead, designing non-uniform or irregular arrays, exploiting the flexibility of antenna arrangement, can enhance the system performance compared to the uniform array. The array can achieve the desired spatial resolution in the directions that matter, and sacrifice other directions. Indeed, a pre-optimized irregular array can be derived, where antenna elements can be arbitrarily deployed within predefined non-overlapping rectangular areas \cite{irshad2025pre}. The antenna arrangement strategy can be ``pre-optimized" before the deployment of the array on-site based on the statistical information with respect to numerous possible user distributions in a particular coverage area. Different from the instantaneous information-based MA technology \cite{10906511} introduced in the next part, this pre-optimized irregular array can be designed based on statistical information with low-complexity but can sometimes achieve about $90\%$ performance for the MA technology. Notably, irregular arrays are particularly useful when the arrays are sparse, where there is room for irregular arrangement. Compared with uniform sparse arrays, there are no grating lobes in irregular arrays, even if there will be large side-lobes.

$\bullet$ \emph{T10: Flexible Widely Spaced Subarrays-Based Technology}

Near-field communication has gathered great interest \cite{ZheSurvey}. One promising potential in spherical curvature-based near-field communication is ``beamfocusing", where the beampattern resembles a spotlight so that the array can distinguish different users in both the distance and angle domain. To achieve such beamfocusing effect, one can continually increase the number of antennas in the arrays, which is undesirable from the practical deployment perspective. One promising flexible array configuration strategy to achieve near-field beamfocusing is proposed in \cite{bjornson2024enabling}, where a large array is divided into smaller subarrays deployed with some meters apart. Note that planar wavefronts can be locally observed at the individual subarrays, but the spherical curvature can be achieved from the perspective of all subarrays. This effect mimics how animals achieve depth perception by using multiple eyes. This flexible widely spaced subarrays-based technology can be naturally combined with the cell-free topology introduced before, or possibly constitute a first step toward that topology since co-deployment of phase-synchronized arrays is easier than distributed deployment.

\subsection{Flexible Real-Time Modifications-Based Technology}\label{Sec_real}

The previously discussed MIMO technologies feature flexible deployment and geometry, but are either optimized once and for all, or modified slowly compared to the channel coherence time. In this part, we discuss flexible MIMO technologies leveraging real-time modifications, driven by instantaneous small-scale channel fading properties at a pace faster than the coherence time (i.e., at the millisecond level). Specifically, these real-time modification–based technologies rely on antenna movements driven by instantaneous channel fading properties, which can be further categorized into movements along the array and movements perpendicular to the array. Note that some of the previously introduced technologies, such as rotatable array technology, can also be adapted based on instantaneous channel fading characteristics. The specific implementation rule can be flexibly chosen based on practical requirements.

$\bullet$ \emph{T11: Antenna Movement Along Array}

For this movement mode, the antenna elements can flexibly move along the antenna array to improve the instantaneous channel conditions. These movements can be discrete or continuous, but within certain confined regions, which are driven by mechanical controllers. Compared with the conventional fixed antenna technology, this antenna movement technology can let the antennas sample the impinging fields at their peaks or nulls to improve the SINRs or channel rank. This technology is also called ``movable antenna (MA)" \cite{10906511} and ``fluid antenna" \cite{10753482}. The antenna locations can be optimized for different metrics, including the rate on one subcarrier or the average rate over many subcarriers. The core difference from the technology in \emph{T9} is that the antenna locations change wherever the active user set changes or there is substantial user movement. In contrast, \emph{T9} optimizes the antenna locations beforehand, to suit an entire statistical user population.

$\bullet$ \emph{T12: Antenna Movement Perpendicular to Array}

In the previous technology, antenna movements are along the array. However, with the aid of micro- and nano-fabrication and flexible metamaterials, the antenna elements can move perpendicular to the array to showcase soft characteristics in the perspective of the whole array. This technology was first studied in the field of EM and the authors in \cite{an2024emerging} applied this technology to wireless communication networks and called it as flexible intelligent metasurfaces (FIMs). Specifically, each antenna element can flexibly move perpendicular to the array within a confined region with the aid of a controller. By doing this, the array can morph its 3D shape in a soft feature. The angles of arrival (AoA) and angles of departure (AoD) can be dynamically adjusted and thus the channel response also varies as the the antenna movement. Indeed, the antenna movement strategy perpendicular to the array can be optimized dynamically based on the instantaneous channel information. 

\subsection{Lessons Learned}\label{lessons}
All these twelve technologies have the potential to be deployed in future wireless networks, which have respective prototypes in different developing stages. Some technologies, such as \emph{T1} and \emph{T2}, are already mature enough to be implemented in practical communication scenarios. Some other technologies, such as \emph{T4}, \emph{T11}, and \emph{T12}, are still in the early stages with some initial prototypes. Beyond these twelve technologies, electronically reconfigurable architectures, such as electronically steerable parasitic array radiators (ESPARs), reconfigurable-pixel antennas, and dynamic metasurface antennas (DMAs), have also been extensively studied. Compared with mechanically reconfigurable systems (e.g., movable/fluid antennas), these electronically reconfigurable architectures achieve compactness, low latency, and high integration capabilities, making them mature and well-suited for near-term deployment. Mechanical approaches provide greater adaptability but depend on motors or actuators, likely resulting in higher cost, power consumption, and slower reconfiguration. Viewed together, electronic and mechanical reconfigurability are complementary: electronic solutions offer practical deployment today, while mechanical solutions explore the ultimate limits of flexibility and inspire future designs, where they define a broader and more balanced landscape for the evolution of flexible MIMO technologies. While mechanical reconfigurability provides valuable insights into flexible MIMO, its scalability and practicality remain an open challenge.

Note that the aim of introducing the terminology ``flexible MIMO" is not to coin another isolated label or rebrand the existing MIMO technologies, but to formulate a useful taxonomy framework for many promising MIMO technologies. ``Flexibility" is an essential property that connects many emerging MIMO technologies, where different flexibilities for MIMO can be explored to formulate an integrated MIMO design framework. For potential readers and practitioners, we want to share the following motivations.

\emph{First}, we motivate the design ideas and provide some insightful research clues as in Fig.~\ref{Clue_Reference}. One intuitive implementation of future flexible MIMO technology can be motivated based on different requirements. One needs to determine ``What network topology should be used?" based on the fundamentals in \emph{T1-T4}. Then, to explore ``What integrated application is tailored for?", one can capture the useful insights from \emph{T5-T7}. Moreover, relying on the focused network topology and integrated applications, one can further consider ``What array geometry feature should be selected?" from the array level based on \emph{T8-T10}. Finally, one can decide ``Should further real-time modification be applied?" to implement the instantaneous information-based modification as in \emph{T11-T12}.

\emph{Second}, since twelve emerging MIMO technologies that provide different kinds of flexibility are studied in this paper, to also facilitate the actionable depth of these technologies, we summarize the major aspects of all of them in Table~\ref{Comparison}. More specifically, we describe the main technical insights for all flexible MIMO technologies, such as the major ideas, representative characteristics, featured capabilities, intractable technical problems, and practical implementation challenges etc., through Table~\ref{Comparison}.

\emph{Third}, we stress that providing a unified quantitative comparison across all twelve flexible MIMO technologies is not feasible at this stage, as they aim to address different challenges, and it is also not our intention. Many technologies are still in early exploration, with heterogeneous hardware assumptions and distinct operating scenarios, which makes a fair and comprehensive evaluation impossible at this stage. Instead, we aim to provide a taxonomy for these technologies with the clue of ``flexibility" and highlight key characteristics in Table~I, design guidelines in Fig.~\ref{Clue_Reference}. We hope MIMO technology can be significantly evolved in all these potential directions.

\emph{Fourth}, the taxonomy framework, relying on three major classifications, does not mean that each MIMO technology should be totally restricted to one particular classification, where a practical system can make use of multiple of these classes. For instance, rotatable array technology can also be implemented based on the real-time instantaneous channel. However, these observations are not a drawback of our taxonomy framework, but showcase that these technologies can be flexibly utilized based on different practical requirements. We also advocate for utilizing the studied MIMO technologies in this paper with a synergistic and flexible vision, instead of limiting one particular technology to one particular classification. Specifically, we provide two case studies that synergistically combine some of the listed technologies promisingly.

\section{Key Enablers for Flexible MIMO Technology}
\begin{figure*}[t]
\setlength{\abovecaptionskip}{-0.1cm}
\centering
\includegraphics[scale=0.2]{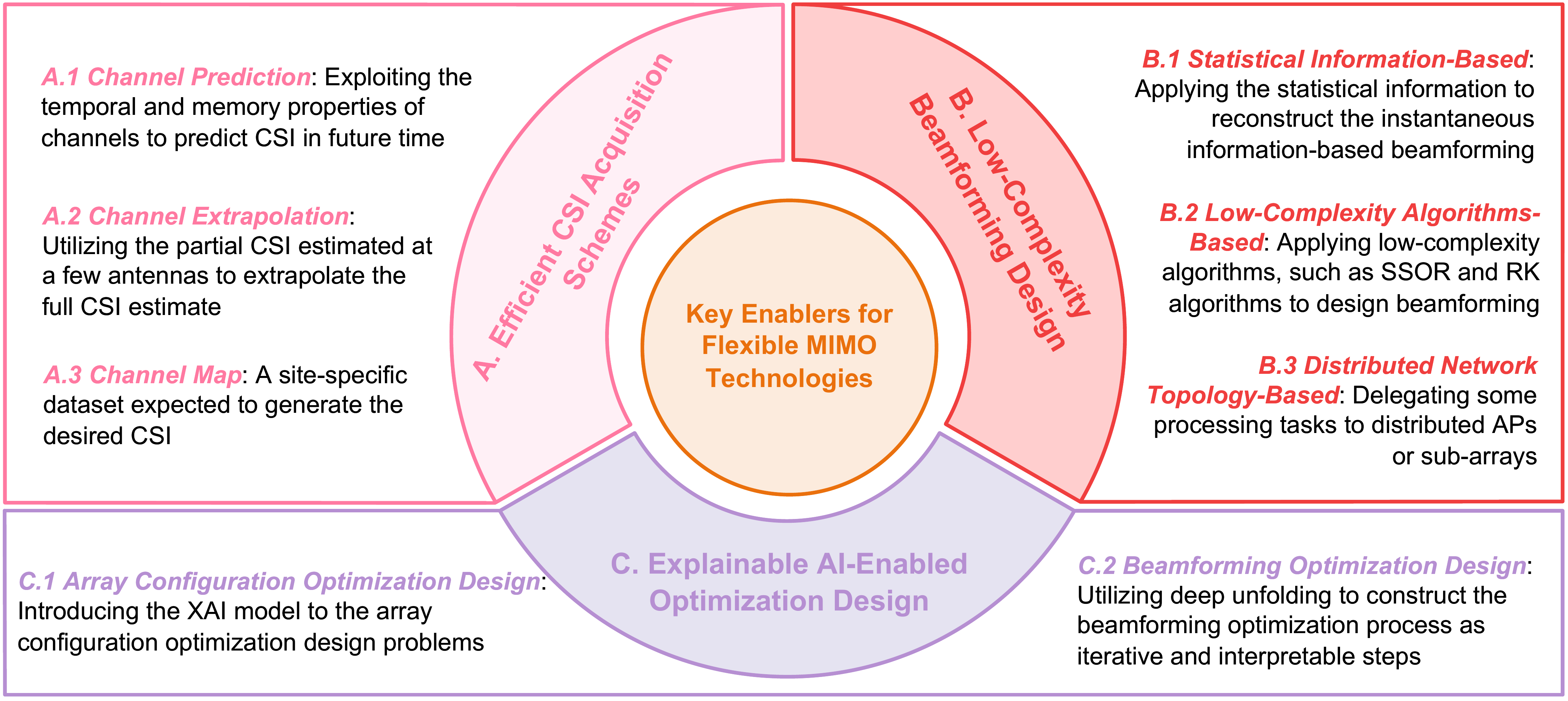}
\caption{Fundamentals of key enablers for flexible MIMO technologies. \label{Enablers}}
\vspace{-0.5cm}
\end{figure*}

Some key enablers can be applied to facilitate these flexible MIMO technologies. The fundamentals of the key enablers in this section are summarized in Fig.~\ref{Enablers}.

\subsection{Efficient CSI Acquisition Schemes}
Note that for a fixed array configuration, one can estimate the channel using classical techniques. Thus, the main issue is to obtain sufficient CSI to also determine desirable array configurations. To efficiently derive the CSI, some promising technologies in different domains can be utilized.

\subsubsection{Channel Prediction}
Channel prediction is a useful technique that focuses on forecasting future channel coefficients utilizing historical and real-time channel coefficients. The basic idea is to exploit the temporal/spatial correlation and memory properties of channels to predict channel coefficients in the future time. To realize it, the predictor antenna can be applied in the mobile communication scenario, such as the high-speed railway scenario \cite{10422885}. Some methodologies, such as the Kalman filter and machine learning, can effectively be applied \cite{10422885}. By applying the time domain-empowered channel prediction, flexible MIMO technologies can proactively adjust the signal processing and resource allocation strategies to mitigate the drawbacks of feedback delay and channel aging.

\subsubsection{Channel Extrapolation}
Channel extrapolation can utilize the partial channel coefficients estimated at a few antennas to extrapolate the full channel coefficients. It can be efficiently applied in flexible MIMO technologies, mainly in two aspects. First, for the antenna array with a large number of antennas, only a few selected antennas need to be activated for the pilot transmission-based channel estimation. Then, the channel coefficients for the other antennas can be efficiently derived based on the space-domain channel extrapolation. Second, for real-time antenna movement-based MIMO technologies, such as \emph{T11} and \emph{T12}, potential antenna positions due to the antenna movement can be regarded as the ``virtual antennas". Thus, the channel coefficients for these virtual antennas can be extrapolated from the channel coefficients for antennas before movement. Note that one cannot perform channel extrapolation in a rich scattering environment because the spatial correlation is weak, but one can do it in sparse environments by estimating parameters of the impinging planar waves \cite{10906511}.

\subsubsection{Channel Map}
The channel map or channel charting is an environment-aware framework, tagged with the locations of transceivers, which is a novel concept in digital twins and can provide location-specific CSI based on inputted geographical locations. The channel map works as a site-specific dataset equipped at stations, which is expected to generate the desired CSI, such as the channel gain, path loss, and even the real-time small-scale fading channel coefficients. Note that the channel map is environment-empowered, which can be efficiently targeted at the specific area. This novel methodology can be efficiently incorporated into flexible MIMO technologies. For instance, relying on the abundant historical channel measurements and accurate radio propagation depiction‌ of the channel map, we can access the potential user distributions and propagation environment features to efficiently pre-design the flexible multi-antenna arrays, such as \emph{T9} and \emph{T11}. More importantly, the channel map methodology can also empower the above three CSI acquisition methodologies with the aid of its prior environment-enabled framework.


One important electromagnetic aspect is mutual coupling, which becomes particularly critical in mechanically reconfigurable architectures. Notably, one main reason to consider sparse arrays (fixed or with movable antennas) is that the mutual coupling becomes small. As antenna positions or orientations change, the mutual coupling among elements varies, altering the effective channels. To efficiently acquire CSI, the joint estimation scheme that can dynamically disentangle the propagation channel response from the coupling effects can be viewed as an important method. This remains a key open research problem at the intersection of electromagnetic theory and channel estimation.

\subsection{Low-Complexity Beamforming Design}
To promote the practical application of flexible MIMO technologies with many antennas, it is crucial to implement low-complexity beamforming schemes since the rapidly changing CSI and flexible array configuration. To facilitate low-complexity beamforming design, some promising ideas or motivations can be utilized. 
\subsubsection{Statistical Information-Based Design}
Compared with rapidly changing small-scale instantaneous information, statistical information remains constant for a long period of time. Thus, it would be promising to implement the statistical information-based design to lower the computational complexity compared to the conventional instantaneous information-based one. \cite{OBETrans} proposes statistical information-cored receive combining schemes, where statistics-based matrices replace the instantaneous information-based matrix inversion part in MMSE combining. It can be observed that approaching SE performance to that of optimal MMSE combining can be achieved by this beamforming scheme for scenarios with low mobility users and severe pilot contamination. This methodology involves a higher offline precomputation load than classical MMSE combining. However, because it relies on statistical information, the overhead is infrequent and acceptable.

\subsubsection{Low-Complexity Algorithms-Based Design}
Another promising methodology to design low-complexity beamforming schemes is motivated from the algorithm-level. Low-complexity algorithms can be applied for the beamforming design. 5G base stations have already implemented CSI-based MMSE-like beamforming. However, focusing on future MIMO technologies with more flexible network and array configurations, it would be insightful to consider some low-complexity algorithms-based beamforming design. Some low-complexity matrix inversion algorithms, such as the symmetric successive over relaxation (SSOR) algorithm, can be applied, which can provide an approximate calculation of the matrix inversion with much lower computational complexity \cite{ZheSurvey}. Moreover, as an efficient methodology to address systems of linear equations (SLEs), the randomized Kaczmarz (RK) algorithm can also be applied to simplify linear beamforming design in flexible MIMO technologies, which can significantly lower the design complexity with excellent achievable rate performance \cite{ZheSurvey}.

\subsubsection{Distributed Network Topology-Based Design}
The low-complexity beamforming schemes can also be designed from the topology-level. More specifically, the distributed network can be efficiently utilized to realize it. For instance, in the cell-free network topology, each AP can first implement distributed processing locally, and then the CPU can implement the final processing. This methodology can alleviate the computational complexity for the fully centralized processing by delegating some processing tasks to the APs. Moreover, for the antenna array with the extremely large antenna number and array aperture, the array can be divided into a few sub-arrays, where each sub-array can implement the beamforming, which can also alleviate the computational complexity. Notably, in practice, for a particular user, only a few APs or sub-arrays need to be activated to provide service to it, which can also effectively balance the achievable performance and computational complexity.

\subsection{Explainable AI-Enabled Optimization Design}
Explainable AI (XAI), also known as white-box AI (WAI), is an emerging paradigm characterized by enhanced transparency and interpretability. Compared with the traditional black-box AI model, the interpretability and mathematical validation of the AI model can be effectively realized in the XAI model. Numerous foundational theories, such as the Bayesian inference, information bottleneck principle, coding rate reduction, and finite-horizon strategies, can be leveraged to underpin the development of XAI models \cite{yang2025whiteboxaimodelfrontier}. By employing these theoretical frameworks, the explainability, reliability, and sustainability of the XAI model can be significantly strengthened. In particular, integrating advanced information-theoretic methods and rigorous mathematical optimization enables the optimization pathways of the XAI model to become transparent and mathematically tractable. Consequently, the XAI methodology can be effectively harnessed to facilitate the optimization design of flexible MIMO technologies.

\subsubsection{Array Configuration Optimization Design}
As discussed above, the geometry characteristics and antenna movement for the array can be flexibly adjusted to provide enhanced performance. The adjustment strategy design can be viewed as the optimized problem. In practice, however, directly solving such optimization problems can be computationally demanding. Therefore, the introduction of XAI-based optimization methods into array configuration optimization emerges as a promising solution. Specifically, by exploiting XAI’s transparent optimization pathways, the inherent complexity of determining optimal array geometry characteristics and antenna movement strategies can be effectively managed. For instance, an XAI model utilizing Bayesian inference can quantify uncertainties associated with antenna placement decisions, thereby guiding robust and interpretable optimization strategies. 

\subsubsection{Beamforming Optimization Design}
For the beamforming optimization design, the XAI model, leveraging deep unfolding techniques, can enhance the adaptability and interpretability of beamforming solutions. Deep unfolding can explicitly construct the optimization process as iterative and interpretable steps, where the gradient descent method is employed to iteratively refine the beamforming scheme based on local channel feedback. Through this structured and transparent optimization pathway, the XAI model effectively balances the desired signals and interference, enabling precise adjustment of the beamforming scheme. Consequently, the resulting beamforming scheme exhibits improved transmission efficiency and robustness, particularly under stringent resource constraints and dynamic channel environments. 
However, XAI entails nontrivial development and deployment costs. Training is often compute- and data-intensive, requiring large, high-quality datasets and incurring substantial training overhead in time and energy. Moreover, the online complexity of some deep-unfolded or white-box models can exceed that of conventional iterative algorithms, underscoring a trade-off among explainability, performance, and real-time feasibility that must be assessed for the target application.

\section{Case Studies}
In this section, we bring two promising case studies for the flexible MIMO technology to showcase its novel advantages and bring some useful insights.

\begin{figure}[t]
\centering
\includegraphics[scale=0.45]{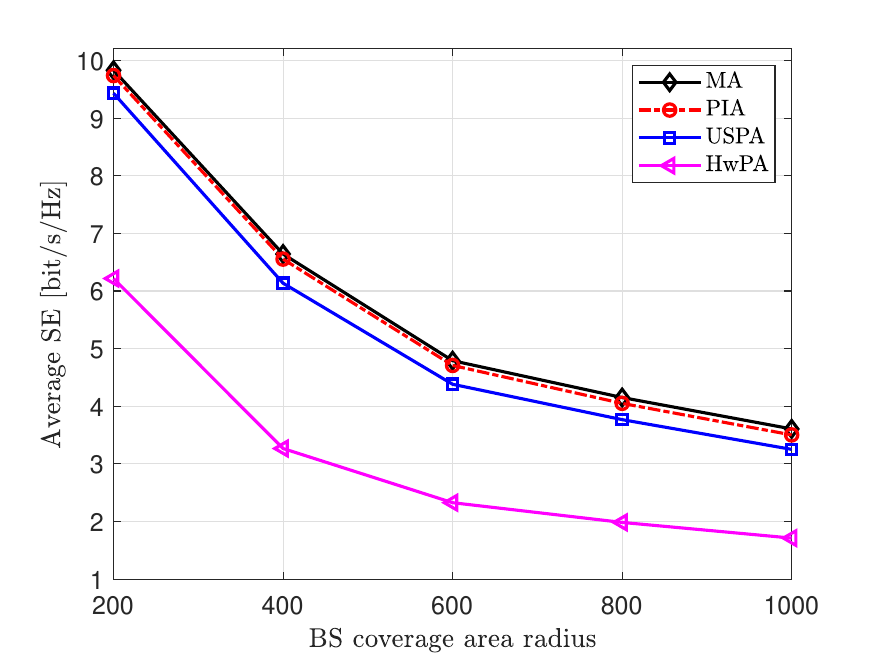}
\vspace{-0.37cm}
\caption{Average SE against the BS coverage area radius for the HSR scenario with $350\,\mathrm{km/h}$ speed. The coverage area radius denotes the distance between the maximum coverage distance and the center of the BS. The BS is equipped with $6\times6$ antennas, serving $8$ single-antenna UEs (carriages) with $28\,\mathrm{m}$ relative height. The location of each BS antenna can be adjusted within a local two-dimensional (2D) region of size $3\lambda \times 3\lambda$ for both the PIA and MA technologies. We sample $80$ realizations of potential positions of carriages along the practical measured railway operation trajectory to optimize the PIA.\label{Sim_HSR}}
\vspace{-0.5cm}
\end{figure}

\subsection{Pre-Optimized Irregular Array-Enabled HSR}
In this part, we utilize the pre-optimized irregular array (PIA) technology introduced in \emph{T9} in the HSR communication introduced in \emph{T7}. As introduced in \emph{T9}, the PIA can be designed based on statistical knowledge of the user distributions within a specific area, which suits the HSR communication scenario with its specific operating area and predictive movement trajectory. We also include the following comparison benchmarks: MA-enabled array, where the array configurations are real-time optimized based on each realization of the user distributions, uniform sparse planar array (USPA) with the BS equipped with a UPA with antenna spacing being $3\lambda$, and half-wavelength planar array (HwPA) with the BS equipped with a UPA with the half-wavelength antenna spacing. Figure~\ref{Sim_HSR} showcases the average SE against the BS coverage area radius for the PIA and the other benchmark schemes. We observe that the PIA outperforms other fixed array configurations, and there is only about $2\%$ gap in average SE between the PIA and the MA-enabled array in the considered setup. This observation demonstrates the potential of synergistically integrating different flexible MIMO technologies.

\begin{figure}[t]
\centering
\includegraphics[scale=0.45]{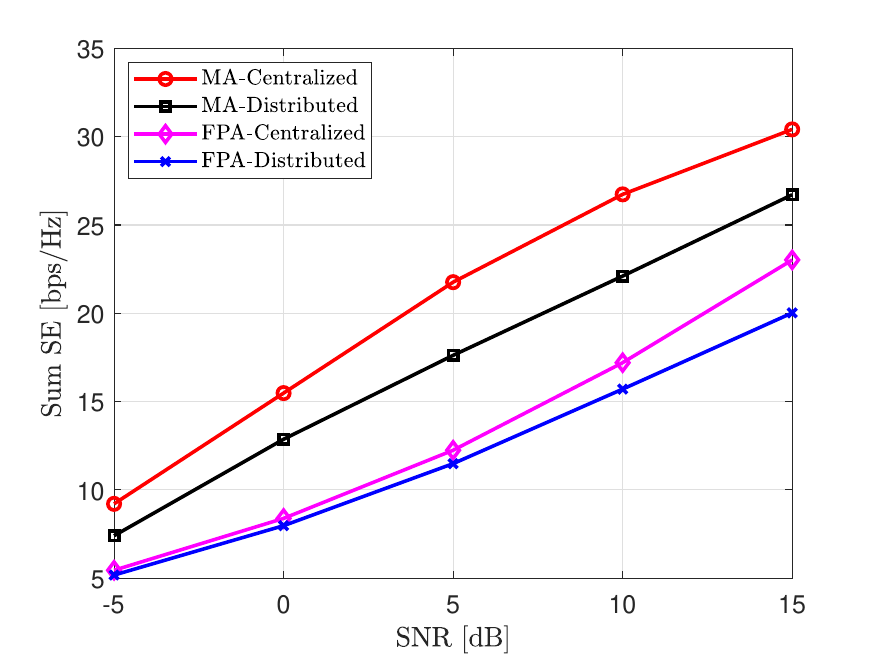}
\vspace{-0.37cm}
\caption{Sum SE for different architectures versus the signal-to-noise ratio (SNR) with $4$ APs and $4$ users. All APs are equipped with UPAs with $2\times2$ isotropic antennas each, and all users are equipped with a single isotropic antenna with $25\,\mathrm{m}$ spacing. The multi-path array response-based channel model without mutual coupling is utilized, and the carrier frequency is $3.1 \, \mathrm{GHz}$. The terminologies ``Centralized" and ``Distributed" denote the centralized and distributed optimization frameworks, respectively. \label{Sim_CF_MA}}
\vspace{-0.5cm}
\end{figure}

\subsection{Cell-Free Movable Antenna}
The cell-free network is a novel topology that can enhance the macro diversity of the system \cite{OBETrans}. The movable antenna technology can enhance the micro diversity of the network by fine-tuning the propagation conditions based on small-scale fading effects \cite{10906511}. In this case study, we combine them and creatively study the cell-free MA technology, where antennas at each distributed AP can flexibly move along the array. The sum SE maximization problem is established to jointly optimize the antenna positions and precoding schemes. The penalty-based block coordinate descent optimization framework, tailored for the cell-free movable antenna network, can be utilized to solve this optimization problem. Fig.~\ref{Sim_CF_MA} investigates the sum SE performance for the cell-free architectures with the MA and FPA. As observed, compared with the conventional FPA-based cell-free architecture, the cell-free MA architecture can achieve much better SE performance, regardless of the centralized and distributed optimization. This case study embraces insightful motivations for the combination of all studied flexible MIMO technologies to facilitate more promising future MIMO technologies. However, although antenna movement can benefit SE, the additional cost, such as motor energy and algorithm complexity, should also be considered, where a trade-off between the performance and cost should be further studied in future.
Meanwhile, our case studies showcase the powerful integration of different flexible MIMO technologies, which demonstrate the synergistic and flexible evolution of future MIMO technologies we advocate in this paper.

\vspace{-0.3cm}

\section{Conclusions}
This paper presented a comprehensive exploration of the flexible MIMO technology. Twelve promising flexible MIMO technologies across three major dimensions, flexible deployment characteristics, flexible geometry characteristics, and flexible real-time modifications, were introduced. Furthermore, we identified three critical enablers: efficient CSI acquisition, low-complexity beamforming, and explainable AI-enabled optimization, along with eight sub-enabling technologies. Through two illustrative case studies on pre-optimized irregular arrays-enabled high-speed railway network and cell-free movable antennas, we validated the substantial capacity enabled by flexible MIMO.

\bibliographystyle{IEEEtran}
\bibliography{IEEEabrv,Ref}

\end{document}